\documentstyle[11pt,aaspp4,natbib]{article}

\newcommand{\vx}{{\bf x}}
\newcommand{\vxp}{{\bf x^\prime}}
\newcommand{\vv}{{\bf v}}

\newcommand{\divg}{{\vec\nabla}\cdot}
\newcommand{\vpred}{{\bf v}_{\mathrm{pred}}}
\newcommand{\vtrue}{{\bf v}_{\mathrm{true}}}

\newcommand{\hmpc}{h^{-1}\mathrm{Mpc}}
\newcommand{\Omegam}{\Omega_{m}}
\newcommand{\Omegal}{\Omega_{\Lambda}}
\newcommand{\betafit}{\beta_{\mathrm{fit}}}

\bibpunct[,]{(}{)}{;}{a}{}{,}
\begin{document}
 
\title{Biased Galaxy Formation And Measurements Of $\beta$}

\author{
Andreas A. Berlind \altaffilmark{1}, 
Vijay K. Narayanan \altaffilmark{2}, and 
David H. Weinberg \altaffilmark{1}
}
\altaffiltext{1}{Department of Astronomy, The Ohio State University, 
Columbus, OH 43210; Email: aberlind,dhw@astronomy.ohio-state.edu}
\altaffiltext{2}{Department of Astrophysical Sciences, Princeton University,
Princeton, NJ 08544-1001; Email: vijay@astro.princeton.edu}

\begin{abstract}

Measurements of the cosmological density parameter $\Omegam$ using techniques
that exploit the gravity-induced motions of galaxies constrain, in linear 
perturbation theory, the degenerate parameter combination $\beta = 
\Omegam^{0.6}/b$, where the linear bias parameter $b$ is the ratio of the 
fluctuation amplitudes of the galaxy and mass distributions. However, the 
relation between the galaxy and mass density fields depends on the complex 
physics of galaxy formation, and it can in general be non-linear, stochastic, 
and perhaps non-local.  The one-parameter linear bias model is almost
certainly oversimplified, which leads to the obvious question: What is the 
quantity $\beta$ that is actually measured by different techniques?  To 
address this question, we estimate $\beta$ from galaxy distributions that are
constructed by applying a variety of locally biased galaxy formation models 
to cosmological N-body simulations.  We compare the values of $\beta$
estimated using three different techniques: a density-density comparison 
similar to the POTENT analysis, a velocity-velocity comparison similar to the 
VELMOD analysis, and an anisotropy analysis of the redshift-space power 
spectrum.  In most cases, we find that $\beta$ estimated using all three 
methods is similar to the asymptotic value of $\Omegam^{0.6}/b_{\sigma}(R)$
at large $R$, where $b_{\sigma}(R)$ is the ratio of rms galaxy fluctuations
to rms mass fluctuations on scale $R$.  Thus, something close to the
conventional interpretation of $\beta$ continues to hold even for complex bias
models.  Moreover, we find that $\beta$ estimates made using these three
methods should, in principle, agree with each other.  It is thus unlikely that
non-linear or scale-dependent bias is responsible for the discrepancies that 
exist among current measurements of $\beta$ from different techniques.

\end{abstract}


\newpage
\section{Introduction}

In cosmological linear perturbation theory, there is a simple relation between
the peculiar velocity and the mass density fields,
\begin{equation}
\vv(\vx) = {f(\Omegam) \over 4\pi} \int \delta_m(\vxp) 
           {(\vxp-\vx) \over |\vxp - \vx|^3} d^3x^\prime ~,
\label{eqn:vlin}
\end{equation}
where $\delta_m(\vx) \equiv \rho(\vx)/\bar{\rho} - 1$ is the mass density 
contrast, $f(\Omegam) \approx \Omegam^{0.6}$, and $\Omegam$ is the ratio of 
the average mass density of the universe to the critical density 
\citep{peebles80}.  The differential form of this relation is
\begin{equation}
\divg\vv(\vx) = - f(\Omegam) \delta_m(\vx) ~.
\label{eqn:divv}
\end{equation}
These relations suggest that if we map the mass density and velocity fields, 
a comparison of the two will yield a measurement of $\Omegam$.  Unfortunately, 
we cannot observe the mass distribution directly.  We can only detect luminous
matter, of which galaxies are the basic unit.  The relation between the galaxy
and mass distributions, usually referred to as ``bias'',  depends on the 
details of the galaxy formation process.  A complete theory of galaxy 
formation should predict what environments galaxies form in and how they are 
distributed in space with respect to the mass.  However, while galaxy 
formation is one of the most actively pursued fields of theoretical cosmology,
our current understanding of it is far from perfect.  The relation between 
galaxies and mass is often parameterized by the linear bias model 
$\delta_g = b \delta_m$, where $\delta_g$ is the galaxy density contrast and 
$b$ is the linear bias factor.  Equation~(\ref{eqn:divv}) then becomes
\begin{equation}
\divg\vv(\vx) = - \beta \delta_g(\vx) ~,
\label{eqn:divvb}
\end{equation}
where $\beta = f(\Omegam)/b$.  

Methods that use equation~(\ref{eqn:divvb}), or some form of it, to infer the
cosmic mass density can only directly measure $\beta$, a degenerate
combination of $\Omegam$ and $b$.  In practice, $\beta$ is usually estimated 
from a comparison between the galaxy density field, inferred from galaxy 
redshift maps, and galaxy peculiar velocities, which require distance 
measurements to individual galaxies.  This estimate is made either through a 
density-density comparison, where the mass density field is predicted from the
smoothed peculiar velocity field and is then compared to the observed galaxy 
density field, or through a velocity-velocity comparison, where galaxy
peculiar velocities are predicted from the galaxy density field and are then 
compared individually to observed galaxy peculiar velocities.  Alternatively, 
$\beta$ can be measured solely from galaxy redshift maps by analyzing the 
anisotropy of galaxy clustering produced by redshift-space distortions, using 
either the power spectrum or correlation function in redshift space.  Based on
these three main approaches, there are many techniques that have been 
developed and used to estimate $\beta$ (see \citealt{strauss95} for a review 
of these techniques).

Since the bias relation depends on the complex process of galaxy formation,
it is probably more complicated than the one-parameter linear bias model.  It 
can, in general, be non-linear, stochastic, and possibly non-local 
\citep{dekel99a}.  Therefore, it is not obvious exactly what information a 
measurement of $\beta$ contains.  Moreover, it is not clear whether the 
different techniques for estimating $\beta$ are measuring the same quantity.  
The primary purpose of this paper is to examine these issues and determine
what can be learned about bias from measurements of $\beta$.  We create 
simulated galaxy distributions with various non-trivial bias prescriptions 
and use them to measure $\beta$ with different existing techniques.  The 
general approach is similar to the one we followed in \citet{narayanan00} 
(hereafter NBW), where we investigated the sensitivity of large-scale 
structure statistics to bias.

An additional motivation for this study is that there are large discrepancies
in current $\beta$ measurements made with different methods.  For example, 
analyses of the IRAS 1.2-Jy galaxy redshift survey \citep{fisher95} have 
yielded $\beta_{IRAS}=0.89\pm0.12$ by the density-density approach 
\citep{sigad98}, $\beta_{IRAS}=0.50\pm0.04$ by the velocity-velocity approach
\citep{willick98}, and $\beta_{IRAS}=0.52\pm0.13$ by the redshift-space 
distortion approach (Cole, Fisher \& Weinberg 1995).  Table~1 summarizes 
current $\beta$ estimates for IRAS-selected samples of galaxies made using 
different techniques.  The estimates of $\beta$ span the range $0.4-1.0$, and 
the most extreme measurements differ from each other at the $\sim 4\sigma$ level.  
Some authors have suggested that these discrepancies could be caused by 
complexity in the bias relation affecting different methods in different ways 
\citep{willick97,sigad98,dekel99a}.  For example, each technique effectively 
probes the bias on a different physical scale, so it is plausible that a 
scale-dependent bias relation could cause such discrepancies.  In this paper 
we examine whether the variation in measured values of $\beta$ can be
naturally explained by complexity of bias.

We employ a variety of simple, non-linear, local bias prescriptions to create
galaxy distributions from the outputs of N-body simulations.  We then measure 
$\beta$ using somewhat idealized versions of three different methods: a 
density-density comparison, a velocity-velocity comparison, and an analysis of 
the anisotropy of the redshift-space power spectrum.  We assume that galaxy
positions and peculiar velocities are known perfectly, and we therefore do not
attempt to model the random and systematic errors that exist in real data.  We
are thus able to isolate the effects of different types of local biasing on 
measurements of $\beta$ without worrying about how these measurements will be 
affected by observational errors.  The systematic errors that exist in real 
observations are, of course, very important, but they are best dealt with in 
the context of specific data sets.  We compare our $\beta$ measurements to two
well-defined functions.  The first is $\beta_{\sigma}(R) = 
\Omegam^{0.6}/b_{\sigma}(R)$.  Here, $b_{\sigma}(R)$ is the bias function 
defined as $b_{\sigma}(R) = \sigma_{g}(R)/\sigma_{m}(R)$, where 
$\sigma_{g}(R)$ and $\sigma_{m}(R)$ are the rms fluctuations of the galaxy 
and mass density fields, smoothed with a top-hat filter of radius $R$. The 
second function is $\beta_P(k) = f(\Omegam)/b_P(k)$, where $b_P(k)$ is the
bias function in Fourier space, defined as $b_P(k) = \sqrt{P_g(k)/P_m(k)}$, 
where $P_g(k)$ and $P_m(k)$ are the power spectra of the galaxy and mass 
distributions.  In NBW we demonstrated that, for local bias models, 
$b_{\sigma}(R)$ and $b_P(k)$ become scale-independent at large scales.
We denote the asymptotic large-scale values of $\beta_{\sigma}(R)$ and 
$\beta_P(k)$ as $\beta_{\sigma}$ and $\beta_{P}$, respectively.

\section{Models}

We have carried out N-body simulations of three different cosmological models,
all based on inflation and cold dark matter (CDM). The first is an $\Omegam=1$,
$h=0.5$ model ($h \equiv H_0/100\;{\rm km}\;{\rm s}^{-1}\;{\rm Mpc}^{-1}$),
with a tilted power spectrum of density fluctuations designed to satisfy both 
COBE and cluster normalization constraints.  The cluster constraint requires 
$\sigma_{8m}\Omegam^{0.6}\approx0.55$ \citep{white93}, where $\sigma_{8m}$ is
the rms amplitude of linear mass density fluctuations in top-hat spheres of 
radius $8h^{-1}$Mpc.  Matching the COBE-DMR constraint and $\sigma_{8m}=0.55$ 
with $h=0.5$ requires an inflationary spectral index $n=0.803$ if one 
incorporates the standard inflationary prediction for gravitational wave 
contributions to the COBE anisotropies (see \citealt{cole97} and references
therein).  The other two models have $\Omegam=0.2$ and $0.4$, with a power 
spectrum shape parameter $\Gamma=0.25$ (in the parameterization of 
\citealt{efstathiou92}) and a cluster-normalized fluctuation amplitude 
$\sigma_{8m}=0.55 \Omegam^{-0.6}$.  These two models are open models with no 
cosmological constant ($\Omegal=0$).  Since $\Omegal$ has a negligible effect
on peculiar velocities at fixed $\Omegam$, our results for the two open models
should also hold for flat-$\Omegal$ cosmologies having the same values of 
$\Omegam$ and the same matter power spectrum.  All simulations were run with 
a particle-mesh (PM) N-body code written by C. Park, which is described and 
tested by \citet{park90}.  Each simulation uses a $400^3$ force mesh to follow 
the gravitational evolution of $200^3$ particles in a periodic cube 
$400 h^{-1}$Mpc on a side, starting at $z=23$ and advancing to $z=0$ in 46
steps of equal expansion factor $a$.  We have run four independent
realizations of each of the above cosmological models, and all of the results 
shown in this paper have been averaged over these four realizations.

The bias between galaxies and mass should ultimately be a prediction of a
theory of galaxy formation.  There are currently three theoretical approaches
to predicting how galaxies are distributed in space with respect to the mass.
Semi-analytic models of galaxy formation identify virialized dark matter halos
in moderate resolution N-body simulations and then populate them with galaxies
using analytic prescriptions (e.g., \citealt{kauffmann99,benson00}); high
resolution N-body simulations resolve individual galaxy-sized dark matter
halos within larger halos (e.g., \citealt{kravtsov99,colin99,moore99});
hydrodynamic simulations follow the evolution of both dark matter and baryons,
including the effects of gas cooling, star formation, and feedback, and
identify galaxies based on the location of cold baryonic lumps (e.g.,
\citealt{cen92,katz92,katz96,blanton99,pearce99}).  All three approaches have 
made much progress in the last few years, but we are still far from a complete
and compelling theory of galaxy formation, and even the most ambitious
applications of these techniques have explored only a small selection of
cosmological models and relatively small simulation volumes.  Rather than
adopt a specific, detailed theory of bias, we employ a simple approach to
modeling the ``generic'' effects of bias on large-scale structure
measurements.  We apply plausible biasing schemes to select galaxy particles 
from the mass distributions and check which measurements of $\beta$ are 
sensitive to the details of the biasing scheme and which are robust, extending
the approach used by NBW (see also \citealt{mann98}).

We create galaxy distributions by applying various local biasing prescriptions to
the mass distributions.  In these bias prescriptions, the probability of a
given mass particle being selected as a galaxy depends on the properties
(density, geometry, or velocity dispersion) of the mass distribution, averaged
in top-hat spheres of radius $4\hmpc$, centered on that particle. 
These prescriptions are also used and described in more detail by 
NBW and \citet{narayanan99}.  In brief, the bias prescriptions are:

(1) Semi-analytic: An empirical bias prescription derived by \citet{narayanan99} 
which characterizes the relation between the galaxy and mass density fields
in the semi-analytic galaxy formation models of \citet{benson00}.  

(2) Sqrt-Exp (Square-root Exponential): A bias prescription in which
$(1+\delta_g) \propto \sqrt{(1+\delta_m)}e^{\alpha(1+\delta_m)}$.
Here, $(1+\delta_g)$ and $(1+\delta_m)$ are the galaxy and mass
overdensities, respectively.  This prescription is intended to be monotonic
(for $\alpha>0$) and non-linear, but it can be tuned to allow galaxies to be 
either biased or anti-biased with respect to the mass, and it yields a 
non-trivial bias relation even when $b_{\sigma}=1$.

(3) Power-law: A bias prescription in which $(1+\delta_g) \propto 
(1+\delta_m)^{\alpha}$.  This is similar to the bias relation suggested
by \citet{cen93}, based on hydrodynamic simulations.

(4) Threshold: A bias prescription in which galaxies do not form below some
mass density threshold, and they form with equal efficiency at densities above
that threshold.

(5) Sigma: A bias prescription in which galaxies are selected with equal
probability from mass particles that have a velocity dispersion $\sigma_v$ 
greater than some threshold value.

(6) Sheet: A bias prescription that selects galaxies in regions where the mass
distribution is planar.  We compute the eigenvalues ($\lambda_3 > \lambda_2 >
\lambda_1$) of the moment of inertia tensor in spheres of radius $4 \hmpc$
around each mass particle and select galaxies to be those mass particles that
have the highest ratio of $\lambda_3/\lambda_1$.

(7) High-$z$: A bias prescription identical to the Power-law bias, except
galaxies are selected based on the mass distribution at redshift $z=3$.
This prescription is intended to produce a biasing relation that is stochastic
on the $4 \hmpc$ scale at $z=0$, though still biased in the mean.

Each of the bias models contains one tunable parameter that is adjusted so
that the resulting galaxy distribution has an rms fluctuation, in top-hat
spheres of radius $12\hmpc$, of $\sigma_{12}\approx0.7$.  The values of 
$b_{\sigma}(12)$ for the galaxy distributions are 1.7, 1.0, and 0.67 for 
$\Omegam=1.0$, 0.4, and 0.2, respectively.  In all the analyses in this paper,
we form the mass and galaxy density fields by cloud-in-cell (CIC) binning the 
particle distributions onto a $200^3$ grid.  We create a volume-weighted, 
smoothed velocity field from the discrete galaxy peculiar velocities using 
the method of \citet{babul94}.  Specifically, we first form the momentum 
field by CIC-binning the momentum of every galaxy onto a $200^{3}$ grid.  We 
smooth this momentum field with a Gaussian filter of radius $R_{1} = R_{s}/2$ 
and divide it by a similarly smoothed density field to form a mass-weighted 
smoothed velocity field.  We then smooth this velocity field with another 
Gaussian filter of radius $R_{2} = (R_{s}^{2} - R_{1}^{2})^{1/2}$, so that the 
effective smoothing radius is $R_s$.  Because the second smoothing dominates
over the first, the final velocity field is volume-weighted rather than 
mass-weighted.

\section{Density-Density Comparison}

Density-density comparisons make direct use of equation~(\ref{eqn:divvb})
to measure $\beta$.  The whole process, however, involves many steps.  
Radial peculiar velocities must first be computed for a sample of galaxies 
that have both redshift and (redshift-independent) distance measurements.  
Due to the large uncertainties present in galaxy distance measurements, 
individual galaxy radial peculiar velocities are poorly known.  The 
resulting radial velocity field must therefore be smoothed over a large 
scale to reduce the velocity errors.  The full 3-dimensional velocity field 
is then inferred from the radial velocity field using the POTENT method 
proposed by \citet{bertschinger89}, under the assumption that galaxy 
velocities trace the gravitational potential field of the underlying mass 
distribution.  Finally, the divergence of this smoothed 3-dimensional 
velocity field yields the linear-theory prediction for the smoothed mass 
density field.  On the other side of the density-density comparison, the 
real space galaxy density field is obtained from a galaxy redshift catalog, 
after correcting the galaxy redshifts for peculiar velocities estimated using
the redshift space density field itself.  The real-space galaxy density 
field is then smoothed on the same scale as the velocity field.  After all 
of these steps are taken, the observed galaxy density field may be directly 
compared to the predicted mass density field.  As shown by 
equation~(\ref{eqn:divvb}), the slope of this comparison yields $\beta$.

In practice, a non-linear generalization of equation~(\ref{eqn:divvb})
is used to determine $\beta$ (e.g., \citealt{dekel93,hudson95,sigad98,dekel99}).  
For example, \citet{sigad98} measure $\beta$ from a comparison of the 
IRAS 1.2-Jy galaxy density field, $\delta_{IRAS}$, to the mass density field 
derived from Mark III peculiar velocities.  They predict the mass density 
field using the non-linear approximation
\begin{equation}
\delta_{POT} = -(1+\epsilon_1)f^{-1}\divg\vv + (1+\epsilon_2)f^{-2}\Delta_2
              + (1+\epsilon_3)f^{-3}\Delta_3 ~,
\label{eqn:ddcor1}
\end{equation}
where $\Delta_2$ and $\Delta_3$ are second and third order terms that
involve sums of double and triple products of partial derivatives of the
velocity field, and the three coefficients $\epsilon_1$, $\epsilon_2$ and
$\epsilon_3$ are empirically determined from a family of CDM N-body 
simulations.  In addition, \citet{sigad98} use a non-linear approximation
to convert the IRAS galaxy redshift map to real space.  Since 
equation~(\ref{eqn:ddcor1}) contains higher powers of $f(\Omegam)$,
it is not possible to measure $\beta$ simply by fitting a line through
the $\delta_{IRAS}$ vs. $\delta_{POT}$ relation.  \citet{sigad98} first 
assume a value for $\Omegam$ and then fit a line through the relation to 
get $b^{-1}$.  Their estimate for $\beta$ is then equal to $f(\Omegam)/b$.
Since the initial choice of $\Omegam$ does not seem to affect
the final $\beta$ measurement, it appears that it is indeed a degenerate
combination of $\Omegam$ and $b$, even in this mildly non-linear 
regime.

Aside from addressing non-linear effects, density-density analyses must
deal with a host of problems that arise due to the imperfect nature of
observational data sets.  In particular, the peculiar velocity errors,
when smoothed, give rise to inhomogeneous Malmquist bias, which is 
difficult to correct for.  Moreover, density-density analyses must deal
with issues such as non-uniform sampling and imperfect survey boundaries 
in both the galaxy redshift, and peculiar velocity data.  Sparse sampling 
of the velocity field, for example, makes it particularly hard to smooth 
the velocity data in an unbiased way.  Much work has been done recently 
to understand and control all of these problems \citep{dekel99}. Current 
density-density comparisons smooth the velocity field with a Gaussian 
filter of radius $12\hmpc$ and yield rather high values of $\beta$ compared 
to other techniques (see Table~1 for a summary of current measurements).  
The analysis of \citet{sigad98} yields $\beta_{IRAS} = 0.89 \pm 0.12$.  

We wish to focus directly on the influence of bias on density-density
comparisons, and we therefore adopt a simpler approach to analyzing our
numerical data sets.  We assume perfect knowledge of both the real-space 
positions and the velocities of galaxies in our simulation volume.  We then 
smooth the galaxy velocity field with a Gaussian filter of some radius $R$ 
and compare its divergence to the galaxy density field smoothed at the same 
scale.  We fit a line of form $-(\divg\vv) = \beta \delta_g + C$ to the 
density-density relation in the region $-0.5<\delta_g<0.5$ in order to 
estimate $\beta$.  Figures 1 and 2 show this procedure when the velocity 
and density fields are smoothed at $12 \hmpc$ for the $\Omegam=1.0$ and $0.2$ 
models.  In each panel we show the mean relation between $-(\divg\vv)$ and 
$\delta_g$ for a given bias prescription, along with its $1\sigma$ scatter,
and the best-fit line to this relation (we have omitted the Sigma and High-$z$ 
bias prescriptions because they produce results that are nearly identical to
the Threshold and Power-law prescriptions, respectively).  The slope
$\betafit$ of the best-fit line is indicated at the lower right corner of each
panel.  Also shown, for purposes of comparison, is the asymptotic large-scale 
value of $\beta_{\sigma}(R)$.  Table~2 summarizes the results for all the
biasing models (including the Sigma and High-$z$ bias models) for each of the 
three cosmological models.

For $\Omegam=1.0$ and 0.4, all the bias models, with the exception of the 
Sqrt-Exp and Sheet models, have best-fit estimates of $\beta$ that are in 
excellent agreement with the asymptotic large-scale value $\beta_{\sigma}$, 
deviating not more than $4\%$ from this value.  The Sqrt-Exp model yields a 
best-fit value that is substantially  higher than $\beta_{\sigma}$, whereas
the Sheet model yields a $\beta$ estimate that is somewhat lower than 
$\beta_{\sigma}$.  However, the strong degree of non-linearity in the 
$-(\divg\vv)$ vs. $\delta_g$ relation for both of these models is striking in 
appearance and can probably be ruled out with existing data (e.g., 
\citealt{sigad98}).  For $\Omegam=0.2$, the situation is slightly different.  
The Semi-analytic and Power-law models yield $\beta$ estimates that 
underestimate $\beta_{\sigma}$ by $\sim 10\%$, whereas the Sqrt-Exp model 
yields a $\beta$ estimate that overestimates $\beta_{\sigma}$ by $8\%$.
Of course, analyses of real data sets must deal with the additional 
complications of non-linearity and statistical biases, but these have been
addressed in the papers cited above and depend in detail on properties of the
data sets themselves.  Our results show that once these challenges are met,
the value of $\beta$ derived from POTENT-like analyses should be close to the
asymptotic value of $\beta_{\sigma}$, for a fairly broad range of assumptions
about the form of biasing.

\section{Velocity-Velocity Comparison}

Velocity-velocity comparisons make use of equation~(\ref{eqn:vlin}) to
measure $\beta$.  Roughly speaking, the observed galaxy redshift distribution 
is used to predict the peculiar velocities of individual galaxies using 
equation~(\ref{eqn:vlin}), with $f(\Omegam)$ replaced by an assumed value for 
$\beta$.  These velocity predictions are then compared directly to peculiar
velocity measurements made using redshift-independent distance measurements.  
The best estimate of $\beta$ is that for which the predicted and measured
galaxy velocities show the best agreement.  

Two distinct velocity-velocity methods have been used in recent measurements
of $\beta$.  The first method, described by \citet{willick97}, requires that 
the galaxy distribution be converted from redshift space to real space 
before it is used to predict peculiar velocities.  This is a slightly 
tricky step, since the peculiar velocities are both the final 
product and a required intermediate ingredient in this process.  Once the 
galaxy positions are corrected for redshift-space distortions, the galaxy
density field must be smoothed in order to suppress the effects of non-linear 
evolution and shot noise.  Equation~(\ref{eqn:vlin}) is then used to compute 
the predicted velocity field from the smoothed galaxy density field, assuming 
a value for $\beta$.  Finally, a maximum likelihood analysis (VELMOD) is 
used to find the value of $\beta$ that produces the best match between the 
predicted and observed galaxy velocities, assuming that an individual galaxy's
velocity is the sum of the linear theory prediction and an uncorrelated
``thermal'' velocity.  In practice, there are many technical difficulties 
involved in this process.  To compute the predicted velocity field, the galaxy
density field must be integrated over all space, as shown by 
equation~(\ref{eqn:vlin}).  Consequently, any systematic problems in the
galaxy redshift data (such as empty regions) will affect the predicted 
velocity field everywhere.  Moreover, the true velocity at any location will 
be affected to some extent by density features that are outside the volume 
sampled by a redshift survey.  \citet{willick97} address this issue by adding 
to the predicted velocity field a quadrupole term that models the tidal field
arising from density features external to the volume probed by the IRAS 1.2-Jy
redshift survey.  The current VELMOD analysis compares the velocities
predicted from the galaxy density field of the IRAS 1.2-Jy redshift survey, 
smoothed with a Gaussian filter of radius $3 \hmpc$, to the peculiar
velocities in the Mark III catalog and yields $\beta_{IRAS} = 0.5 \pm 0.04$ 
\citep{willick98}.

The second velocity-velocity method, described by \citet{nusser94} (hereafter
ND94) and first implemented by Davis, Nusser \& Willick (1996), compares the 
galaxy density field to the observed peculiar velocity field directly in
redshift space.  Another important difference between this method and VELMOD
is that \citet{davis96} do not compare predicted and observed velocities of 
individual galaxies.  Instead, they expand both the predicted velocity field 
that is derived from the redshift-space galaxy density field and the velocity 
field that is measured from redshift-independent distances into a set of orthogonal
modes.  They then perform a mode-by-mode comparison of the two fields and 
determine the value of $\beta$ for which the best match is obtained.  The ND94 
method has been applied to estimate $\beta_{IRAS}$ by combining the 1.2-Jy
redshift survey with the Mark III peculiar velocity sample, yielding
$\beta_{IRAS} = 0.6 \pm 0.2$ \citep{davis96}, with a surface-brightness
fluctuation sample, yielding $\beta_{IRAS} = 0.42^{+0.10}_{-0.06}$ 
\citep{blakeslee99}, and with a type Ia supernova sample, yielding 
$\beta_{IRAS} = 0.4 \pm 0.15$ \citep{riess97}.  Most recently, \citet{nusser00}
have applied this method to the Point Source Catalog redshift survey (PSCz)
and the ENEAR peculiar velocity sample to obtain $\beta_{IRAS} = 0.5 \pm 0.1$.

In order to investigate the effects of bias on velocity-velocity measurements 
of $\beta$, we perform a simplified VELMOD-like analysis.  As in the 
density-density comparisons, we assume perfect knowledge of both the
real-space positions and velocities of galaxies in our simulation volumes.  
We smooth the galaxy density field with a Gaussian filter of radius $R$ and use
equation~(\ref{eqn:vlin}) to predict the velocity field.  We then interpolate 
to galaxy positions to find predicted velocities ($\vpred$) for all the
galaxies.  Finally, we compare these with the true galaxy velocities
($\vtrue$).  The slope of the best-fit line to the $\vtrue$ vs. $\vpred$ 
relation is our measurement of $\beta$.  We do not include a constant offset
term in our fit as we did for the density-density analysis, but it would make
little difference to our results because the $\vtrue$ vs. $\vpred$ relations
are very close to linear.  The offset term is important in the density-density
analysis because of the non-linearity of the $\delta_{\mathrm{true}}$ vs.
$\delta_{\mathrm{pred}}$ relations.  Figures 3 and 4 show this procedure 
when the galaxy density field is smoothed with a Gaussian filter of radius 
$3 \hmpc$ for our $\Omegam=1.0$ and $0.2$ models.  In each panel we show the 
mean relation between $\vtrue$ and $\vpred$ for a given bias prescription,
along with its $1\sigma$ scatter, and the best-fit line to this relation (as 
before, we have omitted the Sigma and High-$z$ bias prescriptions because they 
produce results that are nearly identical to the Threshold and Power-law 
prescriptions, respectively).  The slope $\betafit$ of the best-fit line is 
indicated at the lower right corner of each panel.  Also shown, for purposes
of comparison, is the asymptotic large-scale value of $\beta_{\sigma}(R)$.  
Table~2 summarizes the results for all the biasing models (including the Sigma
and High-$z$ bias models) for each of the three cosmological models.

For all three cosmological models, all the bias models, with the exception of the 
$\Omegam=1.0$ Sheet model, have best-fit estimates of $\beta$ that agree well 
with the asymptotic large-scale value $\beta_{\sigma}$, deviating not more
than $11\%$ from this value.  The Sheet model yields a $\beta$ estimate that
is $33\%$ lower than $\beta_{\sigma}$.  

For smoothing lengths larger than $5 \hmpc$, we discovered a systematic error
that affects $\beta$ estimates made using the VELMOD method, even in the case
where galaxies trace mass.  It arises because a smoothed quantity ($\vpred$) 
is compared to an unsmoothed quantity ($\vtrue$), and the errors in the 
predicted velocities are thus correlated with the predicted velocities 
themselves.  This issue is fully explored in a previous paper
\citep{berlind00}.  The mode-by-mode method of ND94 should not be affected by 
this systematic error because it effectively smooths both the velocity and the
density fields in the same manner.  For a smoothing length of $3 \hmpc$, the
influence on VELMOD analyses is small, at least for the cosmological models
considered here.

\section{Anisotropy Of Redshift-Space Clustering}

Both density-density and velocity-velocity methods require measurements
of galaxy peculiar velocities in addition to redshifts.  Peculiar velocity
measurements always contain large statistical and systematic uncertainties 
because they rely on distance-indicator relations that have substantial
intrinsic scatter as well as uncertain calibrations and environmental 
dependencies.  Moreover, these methods usually require the galaxy density
field in real space and therefore rely on a conversion of the observed 
redshift-space galaxy density field.  There is an entirely different approach 
to estimating $\beta$ that does not suffer from either of these problems,
though it does require very large redshift samples for effective application.
It takes advantage of the fact that line-of-sight distortions in redshift
space are caused by the same velocities that density-density and 
velocity-velocity methods must measure independently.  It is possible to 
estimate $\beta$ simply by analyzing these redshift-space distortions, or, 
more specifically, by measuring the anisotropy of the galaxy clustering
in redshift space.

\citet{kaiser87} showed that, in the linear regime, the redshift-space galaxy 
power spectrum [$P_g^S(k,\mu)$] is related to the real-space galaxy power 
spectrum [$P_g^R(k)$] by
\begin{equation}
P_g^S(k,\mu) = (1 + \beta \mu^2)^2 P_g^R(k),
\label{eqn:pkmu}
\end{equation}
where $\mu$ is the cosine of the angle between the line-of-sight and the
wave-vector ${\bf k}$ of a fluctuation in the galaxy density field.  This 
equation is derived using the plane-parallel approximation, i.e, in the case 
where the volume probed is distant enough that all lines of sight to it are 
effectively parallel.  Equation~(\ref{eqn:pkmu}) reveals that the 
redshift-space power spectrum is anisotropic.  The power spectrum of
fluctuations along the line-of-sight is amplified by the amount $(1 + \beta)^2$
with respect to the power spectrum in real space, whereas the power spectrum 
of fluctuations perpendicular to the line-of-sight is not affected at all.  
The redshift-space power spectrum (\ref{eqn:pkmu}) has been shown to reduce to a
sum of monopole ($l=0$), quadrupole ($l=2$), and hexadecapole ($l=4$) terms,
\begin{equation}
P_g^S(k,\mu) = P^S_0(k) L_0(\mu) + P^S_2(k) L_2(\mu) + P^S_4(k) L_4(\mu),
\label{eqn:expansion}
\end{equation}
where $L_l(\mu)$ are Legendre polynomials.  The first two moments are related
to the real space galaxy power spectrum by
\begin{eqnarray}
P^S_0(k) & = & \left( 1 + \frac{2}{3}\beta + \frac{1}{5}\beta^2 \right) P_g(k),
\nonumber \\
P^S_2(k) & = & \left( \frac{4}{3}\beta + \frac{4}{7}\beta^2 \right) P_g(k)
\label{eqn:moments}
\end{eqnarray}
\citep{cole94}. \citet{hamilton92} derived a similar set of relations for the 
multipole moments of the redshift-space galaxy correlation function.
Equations~(\ref{eqn:expansion}) and (\ref{eqn:moments}) provide us with two 
different ways to measure $\beta$ from the redshift space power spectrum 
$P^S(k,\mu)$ (from now on we will drop the subscript $g$ in the galaxy power 
spectrum); using $P^S(k)/P^R(k)$, the ratio of the angle-averaged
redshift-space power spectrum (monopole) to the real-space power spectrum, or
using $P_2(k)/P_0(k)$, the ratio of the quadrupole and monopole moments of the
redshift-space power spectrum.  These ratios are, in principle, measurable and
in linear theory they are related to $\beta$ by
\begin{eqnarray}
\frac{P^S(k)}{P^R(k)} & = & 1+\frac{2}{3}\beta+\frac{1}{5}\beta^2,
\label{eqn:SR} \\
\frac{P_2(k)}{P_0(k)} & = & \frac{(\frac{4}{3}\beta+\frac{4}{7}\beta^2)}
{(1+\frac{2}{3}\beta+\frac{1}{5}\beta^2)}.
\label{eqn:20}
\end{eqnarray}

These ratios yield an estimate of $\beta$ at each wavenumber $k$.  However,
non-linearity in the velocity and density fields produces distortions in an
opposite sense to the linear theory predictions, leading to systematically 
lower estimates of $\beta$ even on scales as large as $50 \hmpc$.  Hence, it 
is essential to model the non-linearities accurately in order to estimate 
$\beta$ using currently available redshift surveys.  \citet{cole95} estimated 
$\beta$ from $P_2(k)/P_0(k)$ by assuming an exponential velocity distribution 
model, in which galaxies have uncorrelated small scale peculiar velocities
drawn from an exponential distribution in addition to their linear theory 
velocities.  Applying this method to the IRAS 1.2-Jy redshift survey, they 
found $\beta_{IRAS} = 0.52 \pm 0.13$.  \citet{fisher96} modeled the
non-linearity in $P_2(k)/P_0(k)$ by using the Zel'dovich approximation 
\citep{zeldovich70}, thus assuming that the scale dependence in this ratio is 
caused by coherent, rather than random, non-linear motions.  They found 
$\beta_{IRAS}=0.6\pm0.2$.  Finally, \citet{hatton99} proposed and tested an 
empirical model for the non-linearity in $P_2(k)/P_0(k)$ by examining the
scale dependence of this ratio in a large number of N-body simulations
spanning a broad range of cosmological parameters.  This model is more general
than the previous ones because it is based on fully non-linear N-body data.

Equations~(\ref{eqn:SR}) and (\ref{eqn:20}) only hold in the linear regime and
the plane-parallel approximation.  In order to use these relations to measure 
$\beta$, we must measure $P^S(k,\mu)$ in volumes that are both large (so that 
they contain fluctuations large enough to be in the linear regime) and far
away (so that all lines of sight to a single volume are approximately
parallel).  These constraints make it difficult to accurately measure $\beta$ 
from the redshift surveys that exist today.  Ongoing surveys, such as the 
2dF redshift survey and Sloan Digital Sky Survey (SDSS), will be much better
suited to this purpose because they will probe very large volumes, although 
even then non-linearity will be important.  To avoid the need for the distant
observer approximation, Fisher, Scharf \& Lahav (1994) measured $\beta$ from 
the IRAS 1.2-Jy redshift survey by expanding the galaxy redshift-space density
field into spherical harmonics and maximizing the likelihood that an assumed 
real-space galaxy power spectrum would yield that specific set of harmonics.  
The free parameters in this spherical harmonic analysis (SHA) are $\beta$ and 
$\Gamma$, the latter determining the shape of the real-space power spectrum.  
\citet{fisher94} obtained $\beta_{IRAS}=0.94\pm0.17$.  More recently, 
\citet{ballinger00} used the SHA method to measure $\beta_{IRAS}=0.40\pm0.10$ 
for the PSCz, and \citet{hamilton00} developed an optimal way to apply the SHA
method to the same survey and measured $\beta_{IRAS}=0.41^{+0.13}_{-0.12}$. 
Table~1 gives a summary of $\beta$ measurements from redshift-space
distortions in IRAS selected galaxy redshift surveys (also see 
\citealt{hamilton98} for a review of such measurements).

For our simulation analyses, we estimate $\beta$ from the anisotropy of the
redshift-space power spectrum using an approach that is similar to that of
\citet{cole95}, but somewhat simplified.  We take the line-of-sight direction
to be a Cartesian axis of the simulation cube, implicitly assuming that the
whole simulation volume is far enough away to satisfy the distant observer
approximation.  We measure $P^S(k,\mu)$ and $P^R(k)$ using a Fast Fourier
Transform (FFT), and we extract the multipole moments by fitting
equation~(\ref{eqn:expansion}) to $P^S(k,\mu)$.  We compute the average values
and $1\sigma$ uncertainties of the ratios $P^S(k)/P^R(k)$ and $P_2(k)/P_0(k)$
from the four independent realizations of each galaxy distribution.  The
points in Figures 5 and 6 represent the functions $\beta(\lambda)$ obtained
from solving for $\beta$ in equations~(\ref{eqn:SR})and~(\ref{eqn:20}),
respectively, where $\lambda=2\pi/k$.  These functions asymptote to a constant
value only on large scales, since equations~(\ref{eqn:moments}) only hold
under the assumption of linear theory.  We estimate $\beta$ for each bias
model using two methods.  (1) We use the \citet{cole95} exponential velocity
distribution model to model the non-linearity in $P^S(k)/P^R(k)$, and we
estimate $\beta$ by fitting this model to our measured $P^S(k)/P^R(k)$.  We
perform the fit only using modes that correspond to scales $\lambda>20\hmpc$,
and the fitting parameters are $\betafit$ and $\sigma_v$ (the small-scale
velocity dispersion).  The thick lines in Figure 5 represent the resulting fits
for $\Omegam=1.0$. (2) We use the \citet{hatton99} empirical model to model the
non-linearity in $P_2(k)/P_0(k)$, and we estimate $\beta$ by fitting this model
to our measured $P_2(k)/P_0(k)$.  As before, we perform the fit only using
modes that correspond to scales $\lambda>20\hmpc$, and the fitting parameters
are $\betafit$ and $k_{\mathrm{nl}}$ (the wavenumber that corresponds to the
scale at which the ratio $P_2(k)/P_0(k)$ is equal to zero).  The thick lines in
Figure 6 represent the resulting fits for $\Omegam=1.0$.  The thin lines in
Figures 5 and 6 show the function $\beta_P(\lambda)=\Omegam^{0.6}/b_P(\lambda)$,
where $b_P(\lambda)\equiv\sqrt{P_g(k)/P_m(k)}$ at $k=2\pi/\lambda$.  In both
figures, the best-fit estimates $\betafit$ are indicated at the lower right
corner of each panel.  Also shown, for purposes of comparison, are the
asymptotic large-scale values of $\beta_{\sigma}(R)$.  Table~2 summarizes
these values and also includes results for the Sigma and High-$z$ bias models,
which are omitted from Figures 5 and 6, and for the low $\Omegam$ cosmological
models.

For all three cosmological models and all bias models, with the exception of 
the $\Omegam=1.0$ Sqrt-Exp and Sheet models, the best-fit estimates of $\beta$
made using the redshift-to-real-space ratio $P^S(k)/P^R(k)$ underestimate 
$\beta_{\sigma}$ by $10-20\%$.  This underestimate occurs even for the
unbiased (mass) models, indicating that it arises because the exponential 
velocity distribution model for the non-linear behaviour of $P^S(k)/P^R(k)$ 
is not accurate at this level.  \citet{cole95} reached a similar conclusion 
about the accuracy of this non-linear model, in a slightly different manner.
The \citet{hatton99} empirical model does a better job fitting the 
quadrupole-to-monopole ratio $P_2(k)/P_0(k)$.  For $\Omegam=1.0$, all of the 
bias models, with the exception of Sqrt-Exp and Sheet, yield $\beta$ estimates 
that underestimate $\beta_{\sigma}$ by $3-8\%$.  For low $\Omegam$, all of the
bias models, again with the exception of Sqrt-Exp, yield $\beta$ estimates that
agree well with $\beta_{\sigma}$.

\section{Summary}

Figures 7, 8 and 9 summarize all of our results for $\Omegam=1.0$, 0.4, and
0.2, respectively.  Each panel corresponds to a particular biasing
prescription and shows $\beta$ estimates, as a function of smoothing length, 
made using a density-density comparison (circles) and a velocity-velocity 
comparison (squares).  Also shown, at arbitrary scales, are the $\beta$
estimates made from the anisotropy of redshift-space clustering (triangles).
Solid points represent the smoothing scales that correspond to recent
observational estimates of $\beta$: the density-density (POTENT-like) estimate
made with a $12\hmpc$ Gaussian smoothing (solid circle; cf. \citealt{sigad98}),
the velocity-velocity (VELMOD-like) estimate made with a $3\hmpc$ Gaussian 
smoothing (solid square; cf. \citealt{willick97}), and the redshift-space 
anisotropy estimate made using the quadrupole-to-monopole ratio of the
redshift-space power spectrum, $P_2(k)/P_0(k)$, marked at an arbitrary scale 
(solid triangle; cf. \citealt{cole95}).  Also shown, for comparison, is the 
function $\beta_{\sigma}(R)$ (solid line).

These results allow us to answer the question ``what is $\beta$?''.  In most
cases, to a fairly good approximation, the quantity $\beta$ estimated by these
methods is close to the ratio $\Omegam^{0.6}/b_{\sigma}$, where $b_{\sigma}$ 
is the asymptotic, large-scale value of the function $\sigma_g(R)/\sigma_m(R)$.  
This is true for all three methods and for all three cosmological models.
This reassuringly simple answer is just what one might have hoped for.  In our
models, the bias relation in non-linear and bias factors like $b_{\sigma}(R)$
and $b_P(k)$ are scale-dependent on small scales.  Nonetheless, the value of
$\beta$ has a well-defined, intuitively sensible meaning that holds for a wide
range of such models.  Furthermore, measurements of $\beta$ from
density-density comparisons, velocity-velocity comparisons, and analyses of
the anisotropy of redshift-space clustering, should, in principle, yield
consistent results.

There are some exceptions to this simple characterization.  The clearest 
is the Sqrt-Exp bias model, which has a strongly non-linear relation between
galaxy and mass density.  This model exhibits the greatest scale dependence of
$b_{\sigma}(R)$ and $b_P(k)$, and different methods of estimating $\beta$ give
very different results.  However, the strikingly non-linear shape of the 
$-(\divg\vv)$ vs. $\delta_g$ relation for this model can be ruled out by 
current POTENT analysis of observational data \citep{sigad98}, so a bias
relation with such pathological effects is probably unrealistic.  Another 
exception is the Sheet bias model.  Here the different methods yield
measurements of $\beta$ that agree well with each other, but they do not match
the large-scale value $\beta_{\sigma}$.

Our results imply that non-linear local bias is unlikely to account for the 
large discrepancies that exist between current observational estimates of 
$\beta_{IRAS}$ using different techniques.  We have also performed these
analyses for the non-local bias model considered by NBW, where the relation
between galaxy and mass density on the $4\hmpc$ scale is modulated by the
larger scale environment.  Although the model does not have a clearly defined
asymptotic value of $\beta_{\sigma}$, the different $\beta$ estimation methods
nonetheless give results that are consistent with each other.  We therefore
conclude that the discrepancies among current observational estimates of
$\beta$ probably arise from the interaction between the systematic errors in the
observational data sets (the redshift and peculiar velocity catalogs) and
the specific details of the various analysis methods.  If the observational
challenges can be overcome, measurements of $\beta$ by these methods can yield
a useful, physically meaningful quantity: $\Omegam^{0.6}$ divided by a bias 
factor that characterizes the ratio of rms galaxy and mass fluctuations, on
large scales.

\acknowledgments 
We note with great sadness the passing of Jeff Willick, who has been a leader
in this field for more than a decade.  We are grateful to Jeff and to Scott 
Gaudi and Michael Strauss for helpful input and comments.  This work was 
supported by NSF grant AST-9802568.  AAB and VKN were supported by 
Presidential Fellowships from the Graduate School of The Ohio State University
during phases of this project.

\newpage
\bigskip
\bigskip
\begin{center}
\centerline{\small Table~1. Current $\beta$ measurements for IRAS selected galaxies}
\smallskip
\begin{tabular}{llll}
\hline
\hline
\multicolumn{1}{l}{Data} &
\multicolumn{1}{l}{Method} &
\multicolumn{1}{l}{Paper} &
\multicolumn{1}{c}{$\beta$} \\
\hline
IRAS 1.2-Jy + Mark III & POTENT               & \citet{sigad98}     & $0.89 \pm0.12$ \\
                       & VELMOD               & \citet{willick98}   & $0.50 \pm0.04$ \\
                       & ND94                 & \citet{davis96}     & $0.60 \pm0.20$ \\
                       & ROBUST               & \citet{rauzy00}     & $0.60 \pm0.125$ \\
IRAS 1.2-Jy + SNIa     & ND94                 & \citet{riess97}     & $0.40 \pm0.15$ \\
IRAS 1.2-Jy + SBF      & ND94                 & \citet{blakeslee99} & $0.42^{+0.10}_{-0.06}$ \\
IRAS 1.2-Jy            & $P_2(k)/P_0(k)$      & \citet{cole95}      & $0.52 \pm0.13$ \\
                       & $P_2(k)/P_0(k)$      & \citet{fisher96}    & $0.60 \pm0.20$ \\
                       & SHA                  & \citet{fisher94}    & $0.94 \pm0.17$ \\
                       & $\sigma^2_{\parallel}/\sigma^2_{\perp}$ & \citet{bromley97} & $0.80^{+0.40}_{-0.30}$ \\
IRAS QDOT              & $P^S(k)/P^R(k)$      & \citet{peacock94}   & $1.00 \pm0.20$ \\
                       & $P_2(k)/P_0(k)$      & \citet{cole95}      & $0.54 \pm0.30$ \\
PSCz + ENEAR           & ND94                 & \citet{nusser00}    & $0.50 \pm0.10$ \\
PSCz                   & SHA                  & \citet{ballinger00} & $0.40 \pm0.10$ \\
                       & SHA                  & \citet{hamilton00}  & $0.41^{+0.13}_{-0.12}$ \\
\hline
\end{tabular}
\end{center}
\center Note---We have listed only the most recent estimates for a given data
set from each group.

\newpage
\bigskip
\bigskip
\begin{center}
\centerline{\small Table~2. Estimates of $\beta$ from the biased models, using
different techniques}
\smallskip
\begin{tabular}{lccccc}
\hline
\hline
\multicolumn{1}{l}{ } &
\multicolumn{1}{c}{ } &
\multicolumn{1}{c}{POTENT} &
\multicolumn{1}{c}{VELMOD} &
\multicolumn{1}{c}{$P^S(k)/P^R(k)$} &
\multicolumn{1}{c}{$P_2(k)/P_0(k)$} \\
\multicolumn{1}{l}{Model} &
\multicolumn{1}{c}{$\beta_{\sigma}$} &
\multicolumn{1}{c}{$\beta_{\mathrm{est}}$} &
\multicolumn{1}{c}{$\beta_{\mathrm{est}}$} &
\multicolumn{1}{c}{$\beta_{\mathrm{est}}$} &
\multicolumn{1}{c}{$\beta_{\mathrm{est}}$} \\
\hline
$\Omega=1.0$  & & & & & \\
Mass          & 1.00 & 0.92 & 0.94 & 0.80 & 0.93 \\
Semi-analytic & 0.61 & 0.61 & 0.56 & 0.54 & 0.57 \\
Sqrt-Exp.     & 0.62 & 0.84 & 0.69 & 0.64 & 0.98 \\
Power-law     & 0.60 & 0.60 & 0.57 & 0.52 & 0.55 \\
Threshold     & 0.57 & 0.55 & 0.56 & 0.47 & 0.55 \\
Sigma         & 0.54 & 0.53 & 0.56 & 0.46 & 0.52 \\
Sheet         & 0.66 & 0.53 & 0.44 & 0.46 & 0.52 \\
High-$z$      & 0.60 & 0.58 & 0.55 & 0.52 & 0.58 \\
\hline
$\Omega=0.4$  & & & & & \\
Mass          & 0.58 & 0.56 & 0.54 & 0.50 & 0.64 \\
Semi-analytic & 0.57 & 0.55 & 0.54 & 0.49 & 0.59 \\
Sqrt-Exp.     & 0.77 & 0.93 & 0.80 & 0.69 & 0.76 \\
\hline
$\Omega=0.2$  & & & & & \\
Mass          & 0.38 & 0.36 & 0.35 & 0.32 & 0.43 \\
Semi-analytic & 0.52 & 0.46 & 0.48 & 0.41 & 0.51 \\
Sqrt-Exp.     & 0.66 & 0.71 & 0.64 & 0.59 & 0.57 \\
Power-law     & 0.54 & 0.49 & 0.48 & 0.45 & 0.52 \\
\hline
\end{tabular}
\end{center}
\center Note---POTENT results are shown for a $12\hmpc$ Gaussian smoothing.  
VELMOD results are shown for a $3\hmpc$ Gaussian smoothing.  For
$\Omegam=1.0$, all uncertainties are $\sim 0.005$.  For low $\Omegam$, 
uncertainties are $\sim 0.002$, except for Sqrt-Exp models, where they are 
$\sim 0.04$ for $\Omegam=0.4$ and $\sim 0.02$ for $\Omegam=0.2$.

\newpage
\begin{figure}
\centerline{
\epsfxsize=6.0truein
\epsfbox[18 300 590 700]{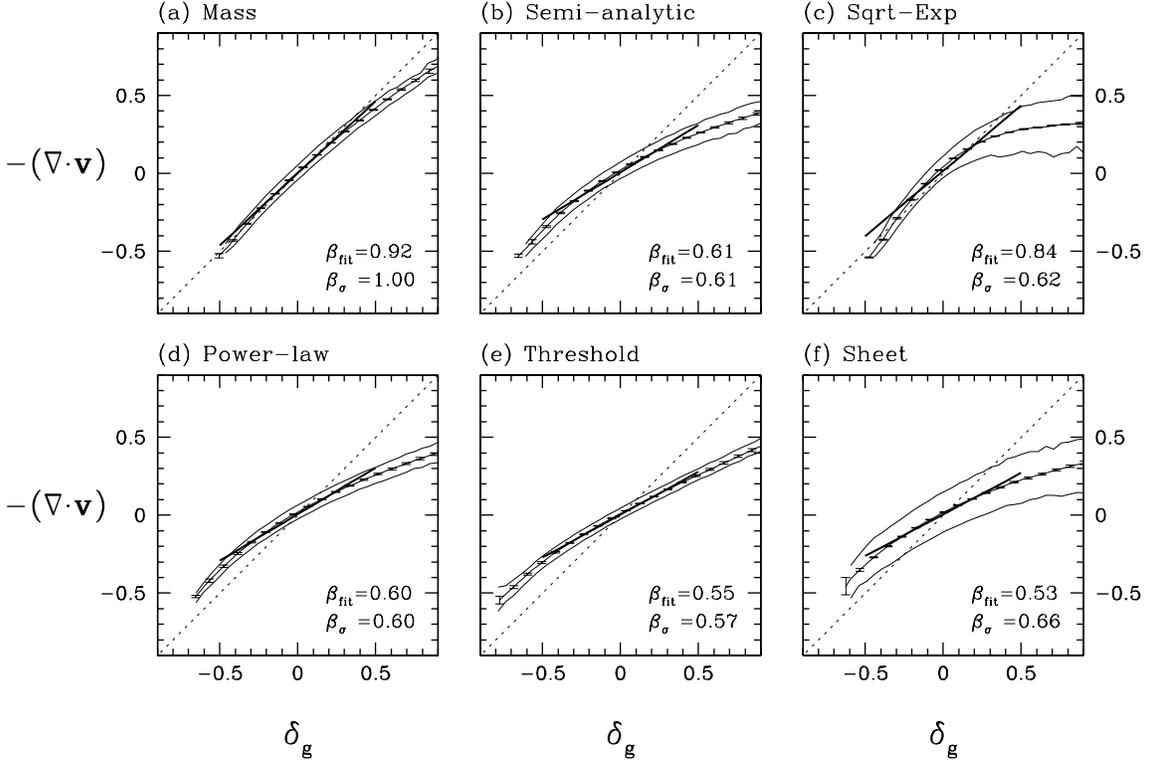}
}
\caption{
Density-density (POTENT-like) $\beta$ estimates for our $\Omegam=1.0$ 
cosmological model.  Each panel shows the mean measured relation between 
$-(\nabla \cdot {\bf v})$ and $\delta_g$ (thin solid curve), when the 
velocity and density fields are smoothed with a Gaussian filter of 
radius $12 \hmpc$, for a particular biasing prescription (we have 
omitted the Sigma and High-$z$ prescriptions because they produce results 
that are identical to the Threshold and Power-law prescriptions, 
respectively).  The plotted relation represents the average over four 
independent simulations, and error bars indicate the $1\sigma$ dispersion
in this mean relation.  The outer two thin solid curves represent the 
locus of points that, for each bin in $\delta_g$, enclose $80\%$ of the 
points in that bin.  Also shown is the best-fit line to the 
$-(\nabla \cdot {\bf v})$ vs. $\delta_g$ relation (thick solid line).  The 
slope of this line, denoted by $\betafit$, is listed in the lower right 
corner of each panel.  Also listed for comparison is the asymptotic, 
large-scale value of $\beta_{\sigma}(R)$.
} 
\end{figure}
\begin{figure}
\centerline{
\epsfxsize=6.0truein
\epsfbox[18 300 591 605]{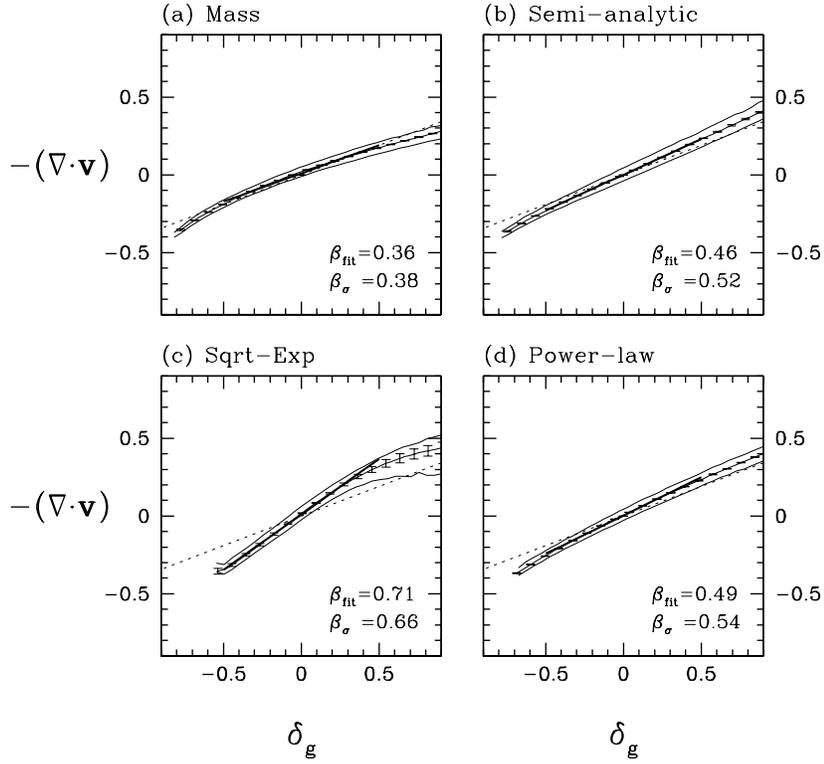}
}
\caption{
Density-density (POTENT-like) $\beta$ estimates for our $\Omegam=0.2$ 
cosmological model. Refer to Fig. 1 for a complete description of the 
components of this figure. 
} 
\end{figure}
\begin{figure}
\centerline{
\epsfxsize=6.0truein
\epsfbox[18 300 590 700]{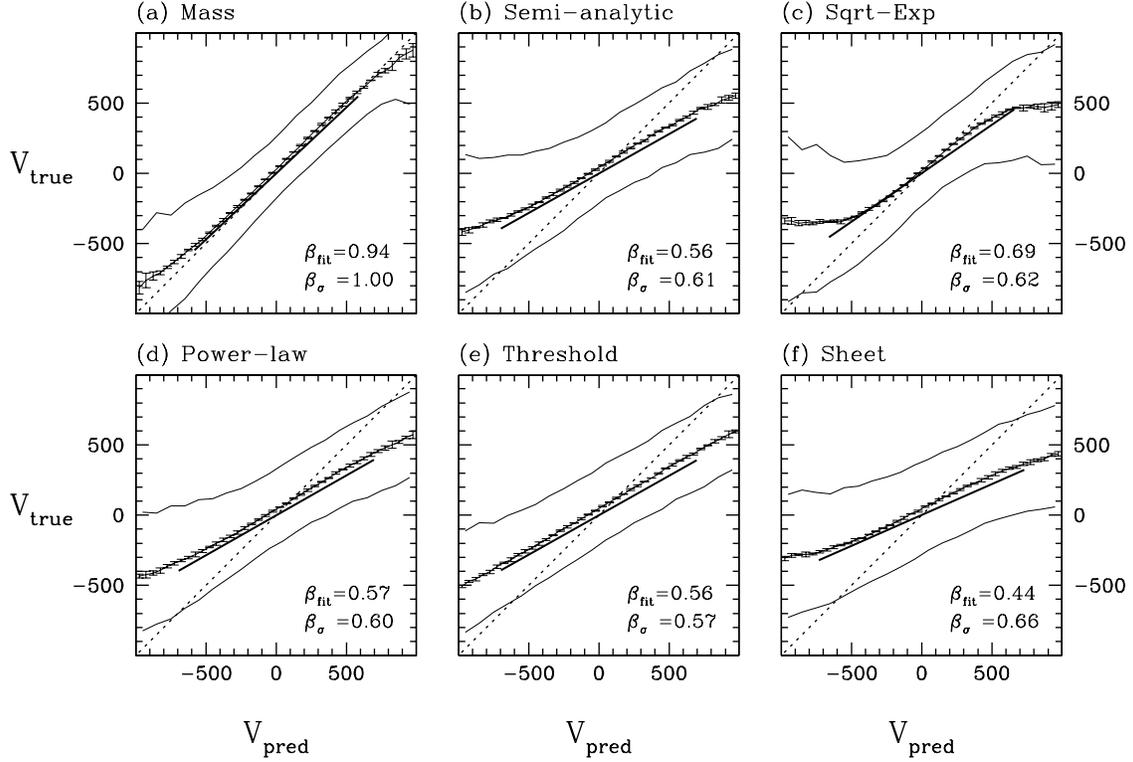}
}
\caption{
Velocity-velocity (VELMOD-like) $\beta$ estimates for our $\Omegam=1.0$ 
cosmological model.  Each panel shows the mean measured relation between 
$\vtrue$ and $\vpred$ (thin solid curve), when the predicted velocities 
are computed from the galaxy density field smoothed with a Gaussian 
filter of radius $3 \hmpc$, for a particular biasing prescription (we 
have omitted the Sigma and High-$z$ prescriptions because they produce 
results that are identical to the Threshold and Power-law prescriptions, 
respectively).  The plotted relation represents the average over four 
independent simulations, and error bars indicate the $1\sigma$ dispersion
in this mean relation.  The outer two thin solid curves represent the 
locus of points that, for each value of $\vpred$, enclose $80\%$ of the 
galaxies.  Also shown is the best-fit line to the $\vtrue$ vs. $\vpred$ 
relation (thick solid line).  The slope of this line, denoted by $\betafit$, 
is listed in the lower right corner of each panel.  Also listed for comparison
is the asymptotic, large-scale value of $\beta_{\sigma}(R)$.
} 
\end{figure}
\begin{figure}
\centerline{
\epsfxsize=6.0truein
\epsfbox[18 300 591 605]{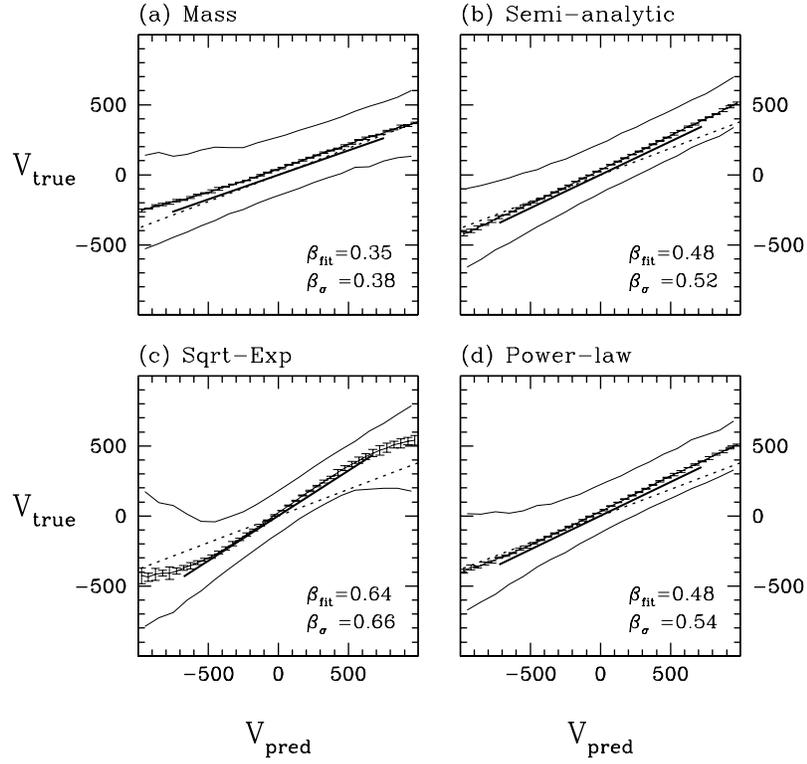}
}
\caption{
Velocity-velocity (VELMOD-like) $\beta$ estimates for our $\Omegam=0.2$ 
cosmological model. Refer to Fig. 3 for a complete description of the 
components of this figure. 
}
\end{figure}
\begin{figure}
\centerline{
\epsfxsize=6.0truein
\epsfbox[18 300 590 700]{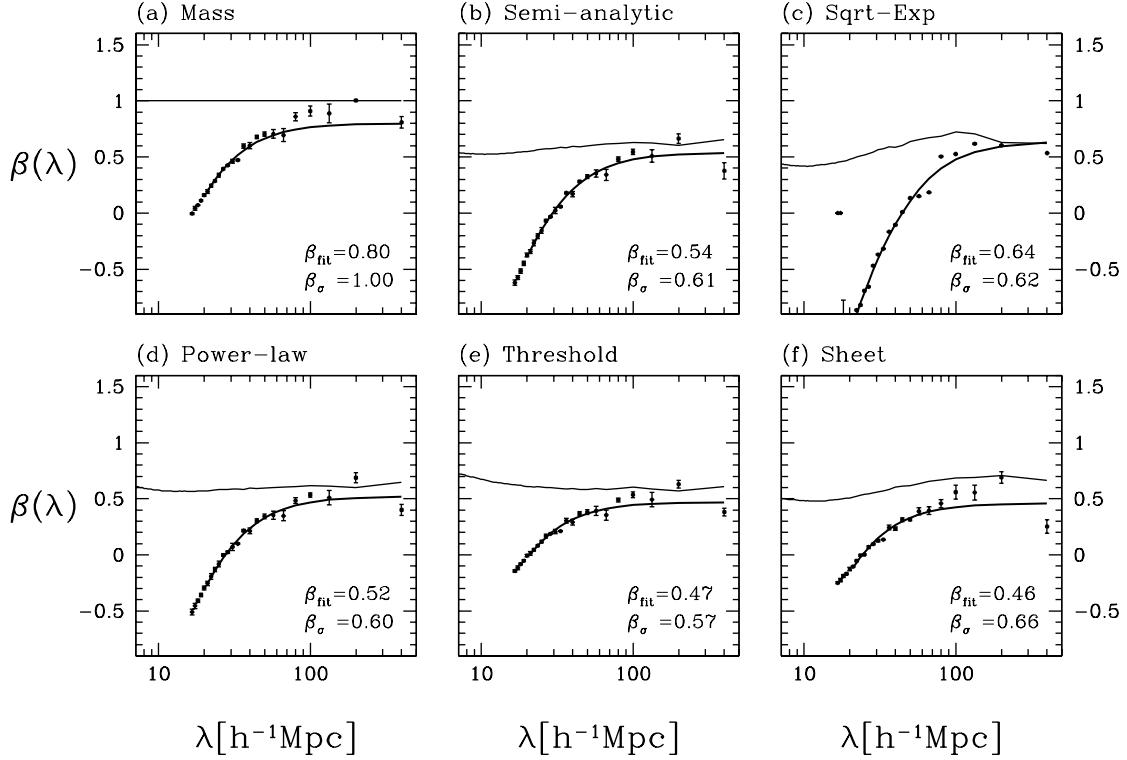}
}
\caption{
Estimates of $\beta$ from $P^S(k)/P^R(k)$, the ratio of the 
redshift- and real-space power spectra, for our $\Omegam=1.0$ cosmological 
model.  Each panel shows an estimate of $\beta$ for a particular biasing 
prescription (we have omitted the Sigma and High-$z$ prescriptions because 
they produce results that are identical to the Threshold and Power-law 
prescriptions, respectively).  The points show $\beta(\lambda)$ as estimated 
from $P^S(k)/P^R(k)$ using linear theory and the plane-parallel approximation
(eq.~\ref{eqn:SR}).  The $1\sigma$ errors in the mean $\beta(\lambda)$ are 
computed using the dispersion among four independent simulations, divided by 
$\sqrt{3}$.  The thin line represents the function $\beta_P(\lambda) =
\Omegam^{0.6}/b_P(\lambda)$, where $b_P(\lambda)$ is the bias function, 
defined as $b_P(k) = \sqrt{P_g(k)/P_m(k)}$, where $P_g(k)$ and $P_m(k)$ are 
the power spectra of the galaxy and mass distributions, respectively.  The 
thick line represents a fit of the exponential velocity distribution model 
to $\beta(\lambda)$.  The fit yields a global estimate of $\betafit$ which 
is listed at the bottom right corner of each panel.  Also listed for 
comparison is the asymptotic, large-scale value of $\beta_{\sigma}(R)$.  
}
\end{figure}
\begin{figure}
\centerline{
\epsfxsize=6.0truein
\epsfbox[18 300 590 700]{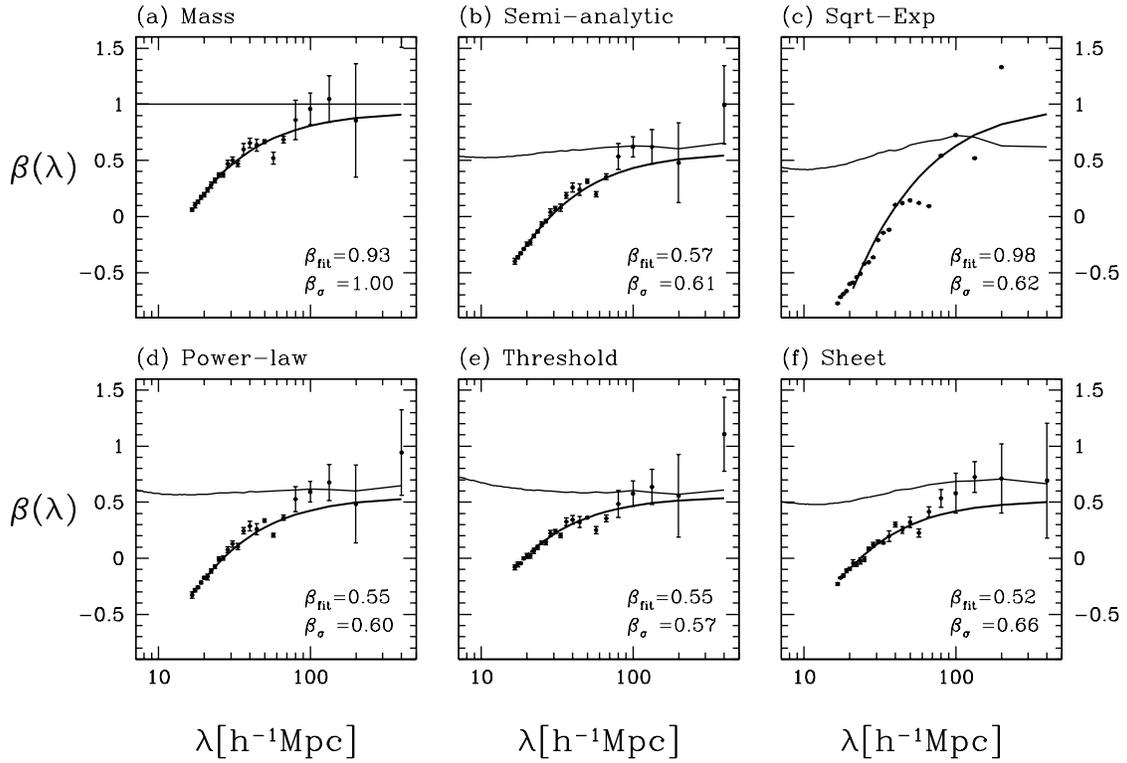}
}
\caption{
Like Fig. 5, except that $\beta(\lambda)$ is estimated via
equation~(\ref{eqn:20}) from $P_2(k)/P_0(k)$, the quadrupole-to-monopole 
ratio of the redshift-space power spectrum, and thick lines represent the
fit of the \citet{hatton99} non-linear model to $\beta(\lambda)$.  
}
\end{figure}
\begin{figure}
\centerline{
\epsfxsize=5.7truein
\epsfbox[18 145 590 695]{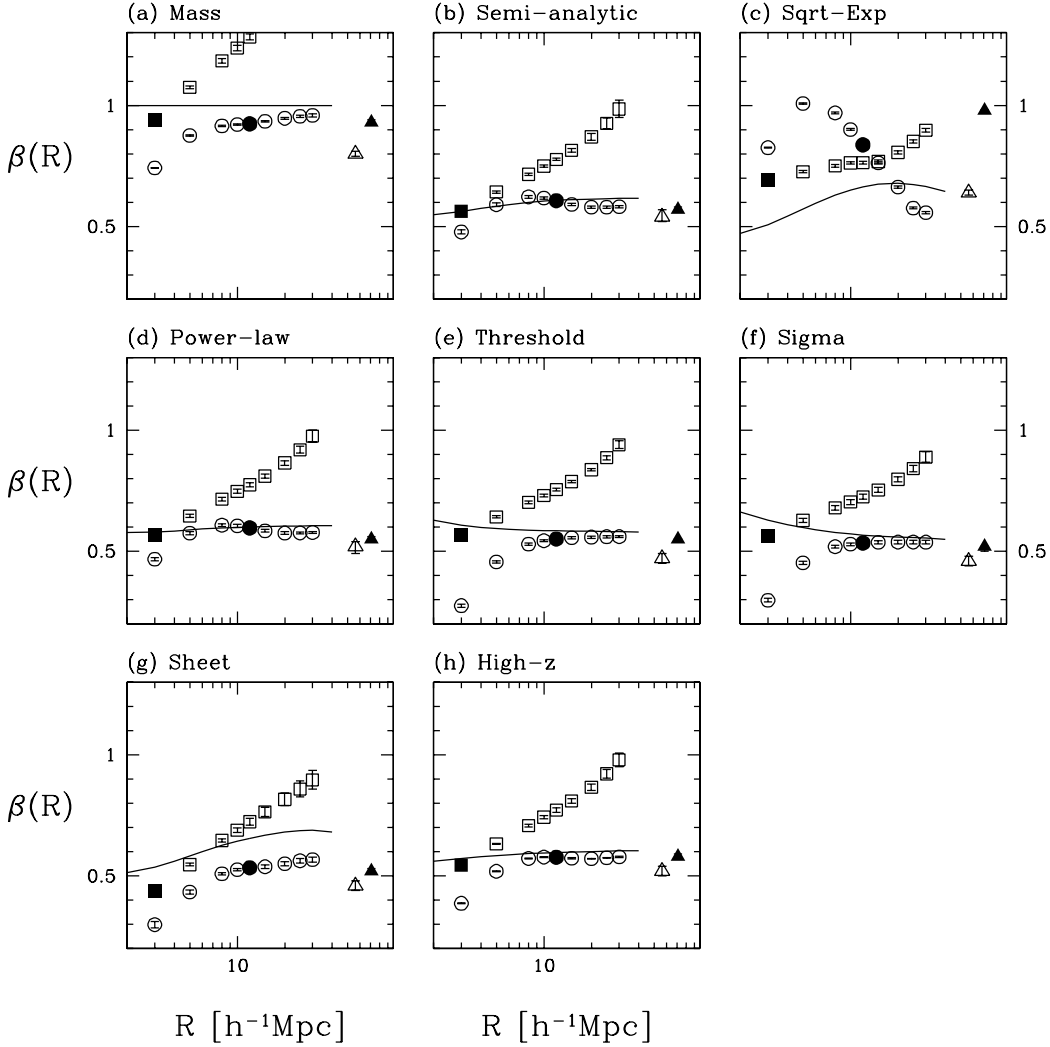}
}
\caption{
Comparison of $\beta$ estimates from different methods, as a function 
of scale, for our $\Omegam=1.0$ cosmological model.  Each panel shows 
$\beta(R)$ for a particular biasing prescription.  The solid line 
represents the function $\beta_{\sigma}(R) = \Omegam^{0.6}/b_{\sigma}(R)$,
with $b_{\sigma}(R) = \sigma_{g}(R)/\sigma_{m}(R)$ for a Gaussian filter 
of radius $R$.  Circles represent density-density (POTENT-like) $\beta$ 
estimates, derived by fitting a line to the measured relation between 
$-(\nabla \cdot {\bf v})$ and $\delta_g$ when the velocity and density 
fields are smoothed with a Gaussian filter of radius $R$ (see Fig. 1).  
The $\beta$ estimate at a $12 \hmpc$ smoothing radius, which corresponds 
to current POTENT measurements, is highlighted as a solid circle.  Squares 
represent velocity-velocity (VELMOD-like) $\beta$ estimates, derived by 
fitting a line to the measured relation between $\vtrue$ and $\vpred$ when 
the predicted velocities $\vpred$ are estimated from the galaxy density field 
smoothed with a Gaussian filter of radius $R$ (see Fig. 3).  The $\beta$ 
estimate at a $3 \hmpc$ smoothing radius, which corresponds to current VELMOD 
measurements, is highlighted as a solid square.  Triangles show estimates of 
$\beta$ derived from the anisotropy of the redshift-space power spectrum. 
These estimates are not scale-dependent and are marked at arbitrary values of 
$R$.  The open triangle shows the estimate of $\beta$ derived from fitting the 
exponential velocity distribution model to $P^S(k)/P^R(k)$ (see Fig. 5).  The 
solid triangle shows the estimate of $\beta$ derived from fitting the 
\citet{hatton99} non-linear model to $P_2(k)/P_0(k)$ (see Fig. 6).  In all 
cases, each point represents the average over four independent simulations, 
and the errorbar represents the $1\sigma$ uncertainty in the mean.  
} 
\end{figure}
\begin{figure}
\centerline{
\epsfxsize=6.0truein
\epsfbox[18 300 591 605]{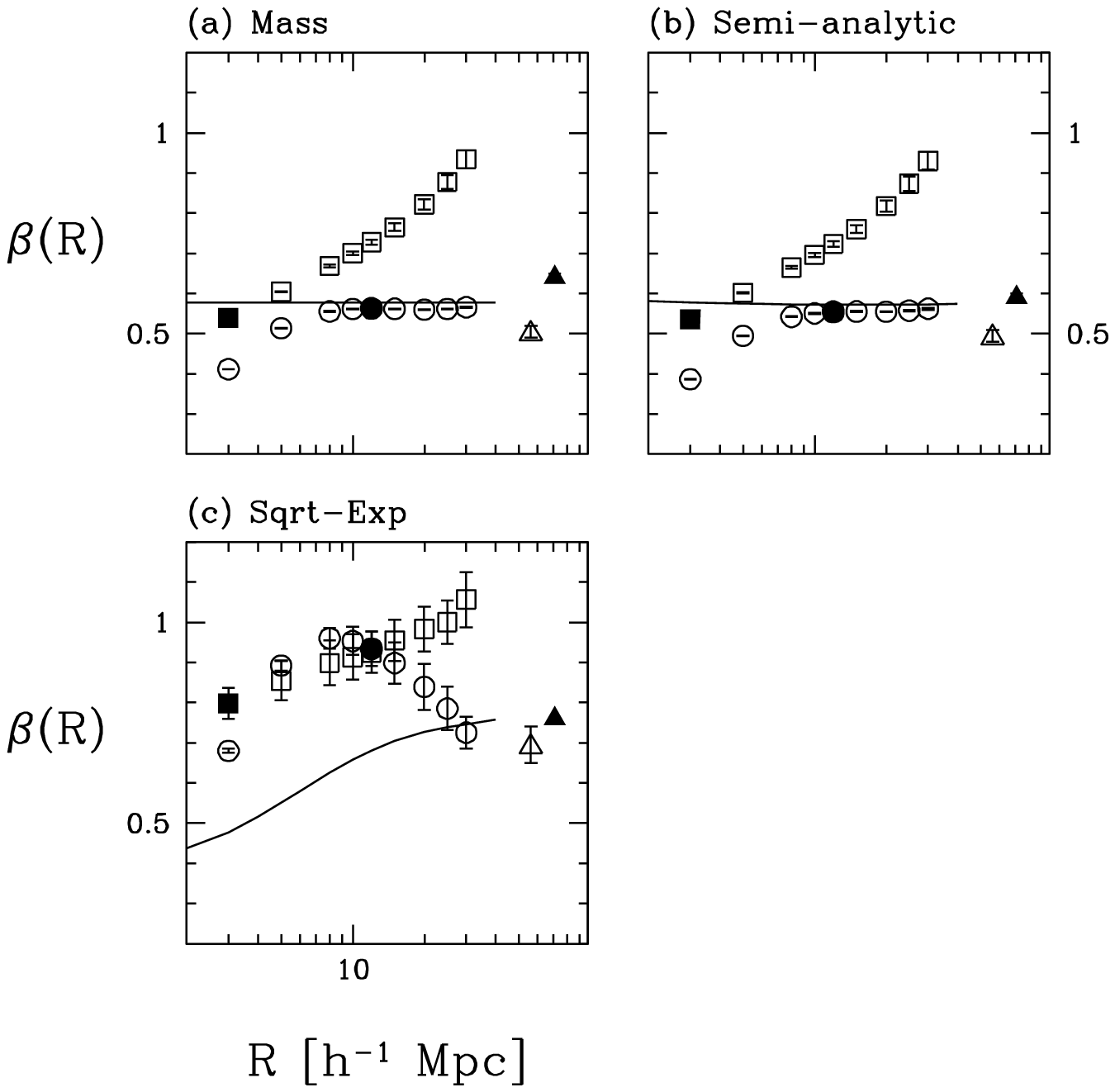}
}
\caption{
Comparison of $\beta$ estimates from different methods, as a function of 
scale, for our $\Omegam=0.4$ cosmological model.  Each panel shows $\beta(R)$ 
for a particular biasing prescription.  Refer to Fig. 7 for a complete
description of the components of this figure.  
} 
\end{figure}
\begin{figure}
\centerline{
\epsfxsize=6.0truein
\epsfbox[18 300 591 705]{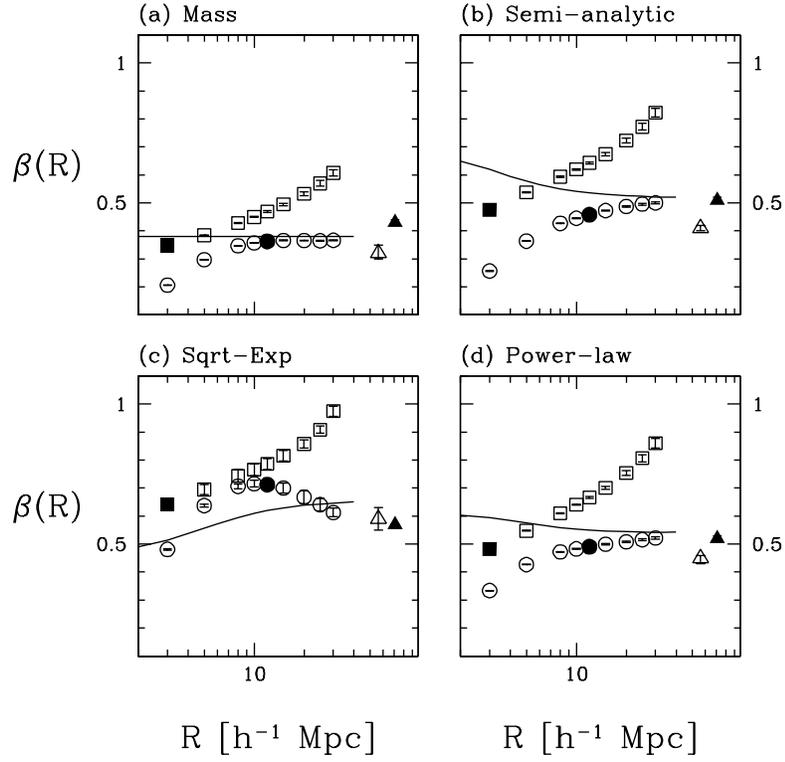}
}
\caption{
Comparison of $\beta$ estimates from different methods, as a function of 
scale, for our $\Omegam=0.2$ cosmological model.  Each panel shows $\beta(R)$ 
for a particular biasing prescription.  Refer to Fig. 7 for a complete 
description of the components of this figure.
} 
\end{figure}
\end{document}